\documentclass[letter]{aa}  
\usepackage{graphicx}
\usepackage{txfonts}

\newcommand{\kms}{km\,s$^{-1}$}

\newcommand{\vsini}{$v_{\rm e} \sin i$}

\newcommand{\lqh}{LQ~Hya}

\newcommand{\figps}[3]{\resizebox{#1}{!}{\rotatebox{#2}{\includegraphics{#3}}}}

\begin{document}

\title{Surface distribution of small-scale magnetic field on the active cool star LQ Hydrae\thanks{Based on observations collected at the ESO La Silla and Paranal observatories (programmes 086.D-0240, 0100.D-0176, 0102.D-0185).}}

\titlerunning{Map of small-scale magnetic field on LQ Hydrae}

\author{
O. Kochukhov\inst{1}
\and T. Hackman\inst{2}
\and J.J. Lehtinen\inst{3,2}
}

\institute{
Department of Physics and Astronomy, Uppsala University, Box 516, S-75120 Uppsala, Sweden\\
\email{oleg.kochukhov@physics.uu.se}
\and
Department of Physics, P.O. Box 64, FI-00014 University of Helsinki, Finland
\and
Finnish Centre for Astronomy with ESO (FINCA), University of Turku, Vesilinnantie 5, FI-20014 University of Turku, Finland
}

\date{Received ; accepted }

\abstract{
It is well known that small-scale magnetism dominates the surface magnetic field topologies of active late-type stars. However, little information is available on the spatial distribution of this key magnetic field component. Here, we take advantage of the recently developed magnetic field diagnostic procedure relying on the magnetic intensification of iron atomic lines in the optical. We extend this methodology from measuring a single average field strength value to simultaneous Doppler imaging reconstruction of the two-dimensional maps of temperature and magnetic field strength. We applied this novel surface mapping approach to two spectroscopic data sets of the young active Sun-like star LQ~Hya. For both epochs, we found a fairly uniform field strength distribution, apart from a latitudinal trend of the field strength increasing from 1.5--2.0 kG at low latitudes to 3.0-3.5~kG, close to the rotational poles. This distribution of the small-scale field does not display a clear correlation with the locations of temperature spots or the global magnetic field structure reconstructed for the same epochs.}


\maketitle

\section{Introduction}

The surface magnetic fields generated by dynamos in late-type stars are fundamentally multi-scale, with structures spanning the entire range of sizes from the stellar radius to photon mean-free path and atmospheric pressure scale height. This is evident both from direct full-disk and local magnetic observations of the Sun \citep[e.g.][]{vidotto:2016,kochukhov:2017b,bellot-rubio:2019} and from theoretical works on numerical simulations of dynamo processes \citep[e.g.][]{yadav:2015,viviani:2018,lehmann:2019}. For unresolved stars other than the Sun, this geometrical complexity of the surface magnetic fields is manifested in the discordant results of magnetic field characterisation using different magnetic field diagnostic methods \citep{donati:2009,reiners:2012,kochukhov:2021}. On the one hand, spectropolarimetric observations, predominantly using high-resolution circular polarised (Stokes $V$) spectral line profiles, are employed to reconstruct detailed maps of global vector magnetic fields with the help of the Zeeman Doppler imaging \citep[ZDI,][]{kochukhov:2016} tomographic inversion procedure \citep[e.g.][]{rosen:2015,rosen:2016,hackman:2016,lehtinen:2022}. Notwithstanding rich information provided by these studies, ZDI analyses miss most of the magnetic flux due to cancellation of opposite field polarities on smaller scales and, as a result, retrieve a grossly underestimated mean magnetic field strength. The former limitation may be one of the reasons why ZDI studies have so far been unable to uncover a systematic spatial relationship between magnetic field and temperature spots, as one would expect from the solar analogy. On the other hand, we can also detect and measure magnetic field by observing Zeeman broadening or intensification of magnetically sensitive spectral lines in the intensity (Stokes $I$) spectra \citep[e.g.][]{robinson:1980,shulyak:2017,lavail:2019,kochukhov:2020}. This approach allows us to determine the total mean magnetic field strength and, in some cases, constrain fractional area coverage (filling factors) of different field strength components, but provides essentially no information on the field geometry. Importantly, due to sensitivity of Zeeman splitting to the absolute value of magnetic field vector, derivation of magnetic information from the intensity spectra is not affected by cancellation of opposite polarities.

The average magnetic field resulting from applications of the ZDI and Zeeman broadening diagnostic methods is occasionally compared in the literature \citep{reiners:2009,vidotto:2014,see:2019,kochukhov:2020,kochukhov:2021}, with the mean field strength derived from the intensity spectra being invariably found much stronger than the average global field calculated from ZDI maps. The interpretation of this discrepancy in terms of the dominant contribution of small-scale fields is undisputed. However, the consequences of this multi-scale nature of stellar magnetic fields for different empirical magnetic diagnostic techniques and contrasting predictions of theoretical dynamo models with observations is yet to be fully understood. 
For example, it is unclear how the additional broadening of polarised spectra by small-scale field may influence their interpretation with ZDI.
It is also unknown if the small-scale magnetic field component exhibits a non-uniform distribution on larger scales, for example, spatially correlating with starspots or global field structures. Furthermore, the impact of line strength and profile variation associated with a sizeable small-scale field on the standard temperature Doppler imaging (DI), which is usually carried out ignoring any magnetic field effects, has not been explored. The goal of this study is to test a new Doppler mapping approach capable of addressing some of these problems. 

The target of the present study is a well-known active star \lqh\ (HD\,82558). It is a single, rapidly rotating \citep[$P_{\rm rot}\approx1.6$~d,][]{jetsu:1993}, young \citep[50~Myr,][]{tetzlaff:2011} K2 dwarf star with parameters similar to those of the Sun at the start of its main-sequence evolution. \lqh\ has a moderately high projected rotational velocity, \vsini\,=\,$28\pm1$~\kms\ \citep{kovari:2004}, enabling an effective DI mapping of starspots and magnetic fields while, at the same time, permitting investigations of magnetic effects on individual line profiles. Surface temperature maps of \lqh\ have been reconstructed in numerous studies \citep{strassmeier:1993,rice:1998,kovari:2004,cole:2015,flores-soriano:2017,cole-kodikara:2019}, revealing a complex, evolving distribution with a mixture of high- and low-latitude spots, but lacking a persistent polar feature commonly found in other active cool stars \citep{strassmeier:2009}. These investigations have typically used detailed radiative transfer calculations of spectral line profiles, while neglecting magnetic effects on model atmospheres and line formation. 

The global magnetic field of \lqh\ was studied with ZDI by \citet{donati:1999b}, \citet{donati:2003} and, most recently, by \citet{lehtinen:2022}. These studies relied on interpretation of the mean Stokes $V$ profiles using a simplified treatment of polarised line formation and approximating the effect of cool spots on spectral lines with a non-uniform continuum surface brightness. The ZDI analyses of \lqh\ found complex global field topologies with a significant toroidal contribution and reported mean surface fields on the order of 150--250~G, with the strongest localised features approaching 1~kG. In parallel, other studies assessed the mean surface magnetic flux of \lqh\ based on intensity spectra. \citet{saar:1996a} measured 2.45~kG field using Zeeman broadening signatures in near-infrared spectra, whereas \citet{kochukhov:2020} determined a field strength of 1.98--2.07~kG at four different epochs from magnetic intensification of optical spectral lines. A discrepancy by one order of magnitude between the mean global field strength obtained from ZDI maps and the total field derived from intensity spectra reinforces the notion that magnetic energy is concentrated on smaller scales not accessible to polarimetry. However, nothing is known about distribution of this dominant field component across the stellar surface since previous Zeeman broadening and intensification studies used single snapshot or time-averaged spectra.

In this pilot study, we extended the Zeeman intensification diagnostic of the small-scale magnetic fields in \lqh\ from measuring a surface-averaged field strength value using a single spectrum to a full spatially resolved Doppler imaging reconstruction of the magnetic field strength map, simultaneously with the temperature distribution using spectral time series. This is similar to the approach proposed by \citet{saar:1992a,saar:1994b}, who attempted to infer a distribution of magnetic flux over the surface of \lqh\ from a set of spectral lines with different Land\'e factors. However, these preliminary studies were inconclusive, yielding inconsistent magnetic field distributions derived from the same data and an average field strength, from 1.02~kG in \citet{saar:1992a} to $\le0.7$~kG in \citet{saar:1994b}; this result is well below all historic and recent Zeeman broadening and intensification measurements of \lqh\ from single spectra. The authors attributed these discrepancies to a cross-talk between magnetic and temperature maps in their inversion procedure and to the impact of a limited signal-to-noise ratio (S/N)  of their observational data. In retrospect, it is also clear that the \ion{Fe}{i} 6173~\AA\ spectral line employed by \citet{saar:1992a,saar:1994b} to extract information on the magnetic field is an inferior choice for a rapidly rotating star compared to the set of \ion{Fe}{i} lines with a strong differential magnetic intensification signature identified by \citet{kochukhov:2020}.

\section{Observations}

The combined temperature and magnetic field mapping analysis applied in this work requires high-resolution time series spectra with characteristics similar to observational data employed for the ordinary temperature DI. For \lqh, several useful data sets can be found in the public archives \citep{lehtinen:2022}. Here we use two sets of observations characterised by the best rotational phase coverage and  high S/N. These are collections of 18 and 9 spectra acquired in February 2011 and December 2017, respectively, with the HARPSpol instrument at the ESO 3.6 m telescope. These data, originally intended for ZDI analysis of the global magnetic field, comprise Stokes $I$ and $V$ spectra covering the 3780--6910~\AA\ wavelength region at the resolving power of $R$\,=\,110\,000. For the purpose of our analysis, we used only the Stokes $I$ observations in the 5430--5510~\AA\ interval, which contains diagnostic lines of interest. The average S/N of \lqh\ spectra in this region is 360 and 410 for the 2011 and 2017 data sets, respectively. The time-averaged HARPSpol spectra from both epochs have been used by \citet{kochukhov:2020} to estimate a mean value of the small-scale magnetic field strength, whereas \citet{lehtinen:2022} have carried out ZDI of the global field topology based on the time-resolved Stokes $V$ profiles extracted from these observations. We refer to these publications for further details of the acquisition and reduction of the HARPSpol observations of \lqh.

\section{Simultaneous temperature and magnetic field mapping}

\subsection{Methodology}

Discerning magnetic field effects on the intensity spectra of cool active stars is a challenging task, particularly for objects with significant rotational broadening such as \lqh. Traditionally, determination of mean magnetic field strength in cool stars was focused on the profile analysis of high Land\'e factor lines at near-infrared wavelengths \citep[e.g.][]{valenti:1995,saar:1996a} or relied on the statistical analysis of a large number of lines \citep[e.g.][]{robinson:1980,lehmann:2015}. In both cases, applications were mostly limited to narrow-line stars and demanded data of very high quality. 
In a recent development, a series of studies demonstrated the possibility
to detect and measure small-scale magnetic fields using differential intensification of spectral lines in the wavelength regions covered by optical spectrographs \citep[e.g.][]{kochukhov:2017c,kochukhov:2019a,kochukhov:2020}. In this case, the magnetic field is measured from line intensities or equivalent widths rather than from line profile shapes, which relaxes the requirements on the spectroscopic data quality and rotational broadening of stellar spectral lines. For this technique to give robust results, a combination of lines with a weak or zero magnetic sensitivity along with lines exhibiting a strong magnetic response is required, but with a similar sensitivity to thermodynamic conditions and element abundances. Ideally, the diagnostic lines should come from the same multiplet, which would allow us to minimise uncertainties due to inaccurate atomic data. Following these principles, \citet{kochukhov:2020} showed that four \ion{Fe}{i} lines ($\lambda$ 5434.52, 5497.52, 5501.47, 5506.78~\AA) from the multiplet 88 \citep{nave:1994} are particularly promising for magnetic studies of G and K-type active stars. In this group of lines, \ion{Fe}{i} 5434.52~\AA\ is not affected by magnetic field, while the remaining three lines show a large magnetic intensification due to a combination of complex Zeeman splitting patterns and relatively large Land\'e factors. Using these lines, \citet{kochukhov:2020} were able to measure 1.5--2.0~kG fields in several active stars with \vsini\,=\,15--30~\kms, including \lqh. \citet{hahlin:2021} have successfully applied the same methodology to the components of the eclipsing binary UV~Psc, which have \vsini\ values 70 and 50~\kms.

\begin{figure}
\centering
\figps{0.9\hsize}{0}{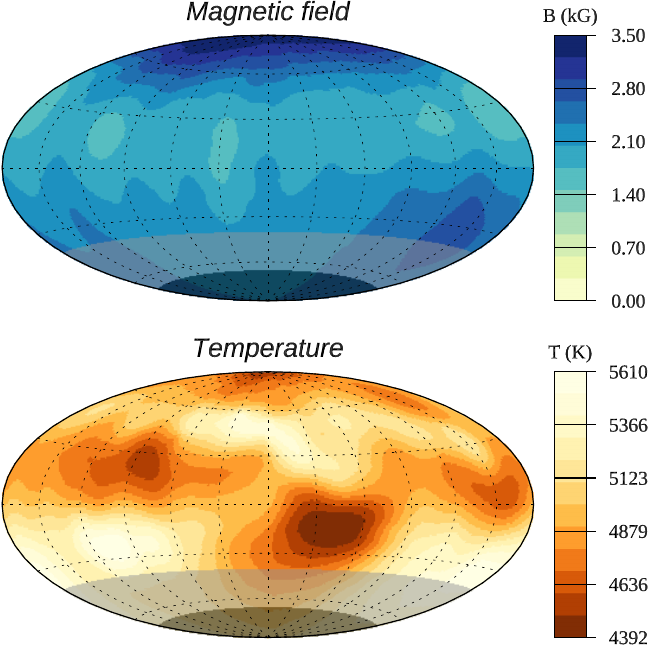}
\caption{Distribution of magnetic field strength (upper panel) and temperature (lower panel) derived from the 2011 data set. The dark grey region shows invisible part of the stellar surface. The light grey highlights part of the surface with a relative visibility below 25\%. The longitude increases left to right, with the central meridian corresponding to 180\degr.}
\label{fig:map2011}
\end{figure}

Here, we implemented the magnetic field diagnostic using Zeeman intensification in a joint temperature-magnetic field DI reconstruction procedure. For this purpose, we employed the {\sc Invers13} ZDI code \citep{kochukhov:2012,kochukhov:2013,rosen:2015}, which is capable of reconstructing maps of vector magnetic field and temperature from the intensity and polarisation profiles of individual spectral lines. The code performs detailed polarised radiative transfer calculations, employing {\sc MARCS} model atmospheres \citep{gustafsson:2008} with different $T_{\rm eff}$ to approximate the quiet photosphere and temperature spots on cool active stars. Since we are interested in characterising distribution of the small-scale magnetic field, which does not contribute to polarisation spectra, we use Stokes $I$ observations alone for the purposes of this study. The temperature and magnetic field maps are recovered from the set of \ion{Fe}{i} lines described above, with information on temperature inhomogeneities obtained from the common distortions of line profiles and local magnetic field strength inferred from the intensification of magnetically sensitive lines relative to \ion{Fe}{i} 5434.52~\AA. The Tikhonov regularisation \citep[e.g.][]{piskunov:2002a} is applied to both maps to ensure uniqueness of the solution. 

The local line profiles are computed by {\sc Invers13} as a function of temperature and small-scale magnetic field. The latter is treated using a radial field, which is a common approach of the Stokes $I$ cool-star magnetic field studies \citep[e.g.][]{marcy:1982,shulyak:2010c,kochukhov:2021}, justified by the lack of sensitivity of intensity profiles to the field orientation. In reality, the local magnetic field strength reconstructed by our code corresponds to a weighted average of multiple unresolved field strength components. With the chosen set of optical diagnostic lines, these components cannot be constrained individually for stars with significant rotational broadening, even when we are using the simplest two-component model \citep{kochukhov:2020}.

\begin{figure}
\centering
\figps{\hsize}{0}{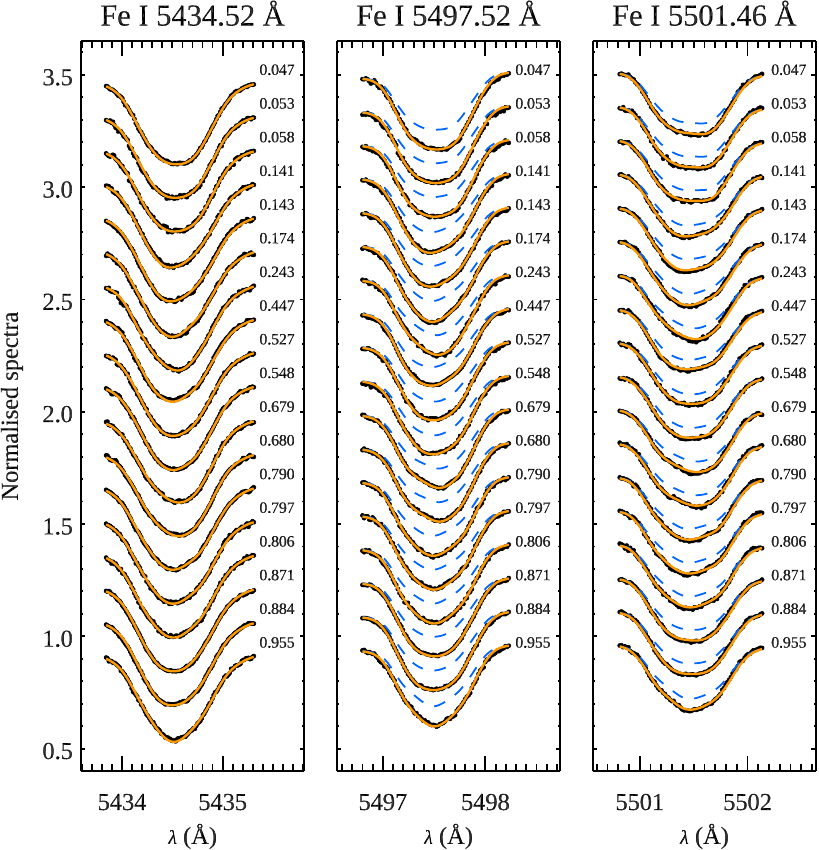}
\caption{Comparison of the observed (black symbols) and best-fitting model profiles (orange solid lines) for the \ion{Fe}{i} 5434, 5497, and 5501~\AA\ lines in the 2011 data set. The blue dashed lines correspond to the synthetic spectrum computed with the same temperature distribution but without magnetic field. The spectra for consecutive phases are shifted vertically, with phases indicated to the right of each profile.}
\label{fig:prf2011}
\end{figure}

The atomic line parameters required for magnetic spectrum synthesis calculations in {\sc Invers13} were obtained from the VALD3 database \citep{ryabchikova:2015}. These data were complemented by a molecular line list complied by \citet{heiter:2021}, with the C$_2$ lines being the most significant molecular absorber in the studied wavelength region. All parameters of the four main \ion{Fe}{i} diagnostic lines, including the improved van der Waals damping constants \citep{barklem:2000b}, were kept fixed, while the oscillator strengths and, in a few cases, wavelengths of other transitions within $\pm1.4$~\AA\ of the main \ion{Fe}{i} lines were adjusted based on comparison with the spectra of benchmark stars. For the latter, we simultaneously fitted the solar flux atlas \citep{wallace:2011} and the $R=220\,000$ ESRESSO spectrum of $\delta$~Eri \citep{adibekyan:2020} using stellar parameters from \citet{heiter:2015}. Despite these corrections and similar to the problems reported by \citet{kochukhov:2020}, the blue wing of the \ion{Fe}{i} 5506.78~\AA\ line, heavily blended by the \ion{Fe}{i} 5505.88~\AA\ and hyperfine-split \ion{Mn}{i} 5505.87~\AA\ features in the rotationally broadened spectrum of \lqh, could not be adequately described by our calculations. Consequently, the \ion{Fe}{i} 5506.78~\AA\ line was omitted, leaving the two \ion{Fe}{i} lines with a strong magnetic response ($\lambda$ 5497.52 and 5501.46~\AA) and one magnetically insensitive \ion{Fe}{i} line ($\lambda$ 5434.52~\AA) for the DI analysis.

For the DI inversion with {\sc Invers13} we adopted the same stellar parameters ($\log g=4.0$, $i=65\degr$, $\xi_t=0.5$~\kms, $\zeta_t=1.5$~\kms, $[M/H]=0.0$) as frequently used by non-magnetic DI studies of \lqh\ \citep[e.g.][]{kovari:2004,cole:2015}. Observations were phased with the same ephemeris as in \citet{lehtinen:2022}, who used $P_{\rm rot}=1.601136$~d, originally from \citet{jetsu:1993}. All inversions were initiated using a uniform temperature distribution with $T=5000$~K and a homogeneous 2.0~kG radial magnetic field according to the results by \citet{kochukhov:2020}. The projected rotational velocity, \vsini, was determined by optimising the fit to observations of the three \ion{Fe}{i} lines while also keeping low-latitude axisymmetric structures in the temperature map to a minimum. The resulting value, $27.5\pm0.5$~\kms, is within the 26.5--28.0~\kms\ range obtained by previous DI studies \citep{donati:1999b,kovari:2004}.

\subsection{Results for 2011 and 2017 data sets}

\begin{figure}
\centering
\figps{0.9\hsize}{0}{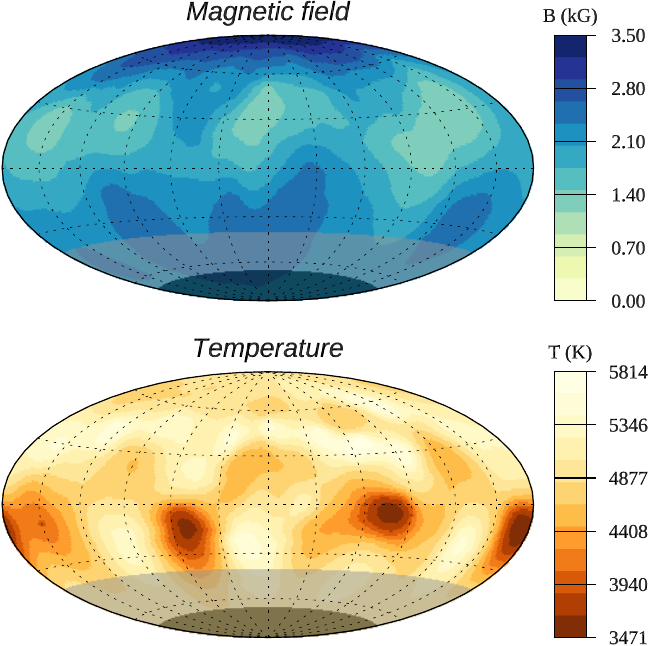}
\caption{Same as Fig.~\ref{fig:map2011} but for the 2017 data set.}
\label{fig:map2017}
\end{figure}

\begin{figure}
\centering
\figps{\hsize}{0}{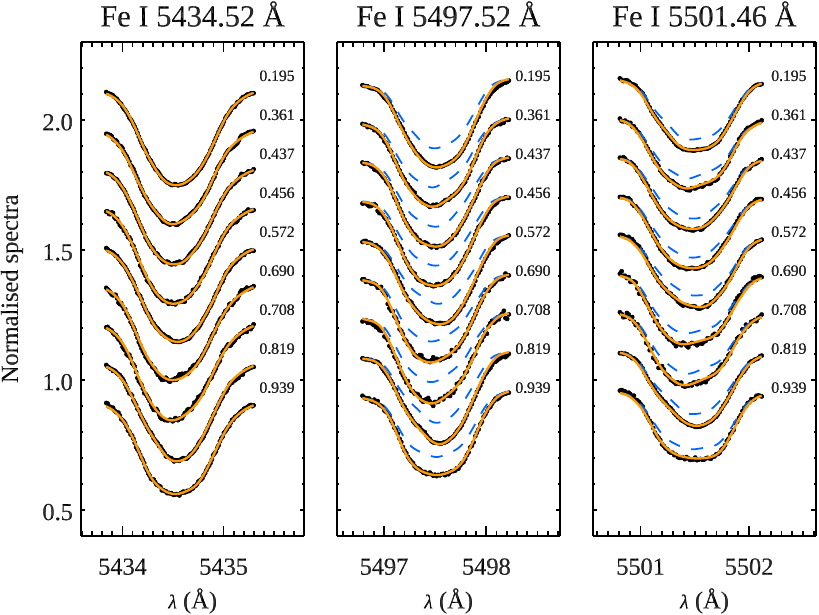}
\caption{Same as Fig.~\ref{fig:prf2011} but for the 2017 data set.}
\label{fig:prf2017}
\end{figure}

The results of simultaneous mapping of magnetic field and temperature spots on the surface of \lqh\ are presented in Fig.~\ref{fig:map2011} for the 2011 data set. The corresponding comparison of the observed and calculated \ion{Fe}{i} line profiles is shown in Fig.~\ref{fig:prf2011}. Furthermore, Figs.~\ref{fig:map2017} and \ref{fig:prf2017} contain the same information for the DI reconstruction applied to the 2017 data set.

The outcome of our analysis for both epochs points to a low-contrast, nearly featureless distribution of the small-scale magnetic field strength. The only departure from this homogeneity is an increase of the field strength towards rotational poles. We have verified that this latitudinal trend is required by the data since replacing it with a single field strength value  adjusted to obtain the optimal fit quality results in a significantly higher $\chi^2$ (up to 40\% increase for the magnetically sensitive lines). The obtained field strength distribution is clearly distinct from temperature maps, for which we derive a lower contrast structure with a single low-latitude feature in 2011 and three higher contrast equatorial cool spots in 2017.

The line profile fits presented in Figs.~\ref{fig:prf2011} and \ref{fig:prf2017} confirm that a strong, multi-kG magnetic field is required to reproduce the observed intensification of the \ion{Fe}{i} 5497.52 and 5501.46~\AA\ lines relative to the magnetically insensitive \ion{Fe}{i} 5434.52~\AA\ line. Moreover, this intensification does not appear to change with rotational phase,
which is consistent with the lack of non-axisymmetric structures in the magnetic field map.

\subsection{Reconstruction from simulated data}
\label{sec:test}

To assess the reliability of the DI reconstruction of temperature and magnetic field distributions using {\sc Invers13}, we carried out a numerical experiment based on simulated observations. The temperature map reconstructed for the first epoch was combined with a simple field distribution comprising a 3 kG circular feature, centred at the latitude $+30\degr$, superimposed onto a 1 kG field background (Fig.~\ref{fig:test}a). We calculated the Stokes $I$ profiles of the three \ion{Fe}{i} lines for the same set of rotational phases as in the observations from 2011 and added a random Gaussian noise with the standard deviation matching that of individual observations. We then recovered $T$ and $B$ maps with the same initial guesses and regularisation as applied in the analysis of real observations.

The resulting maps are shown in Fig.~\ref{fig:test}b with the corresponding line profile fits in Fig.~\ref{fig:test}c. This test demonstrates that a magnetic field structure with a large contrast can be successfully reconstructed with our technique without a significant cross-talk with the temperature map. Both the latitudinal and longitudinal positions of the magnetic spot are correctly reproduced, although the spot is somewhat smeared in the latitudinal direction. The input and reconstructed temperature maps are very similar. Quantifying the difference between the maps in Figs.~\ref{fig:test}a and b, we find a standard deviation of 78 K and 332 G for temperature and magnetic field, respectively. 

In a further series of tests (not shown here), we explored reconstruction of the small-scale field strength for situations when magnetic concentrations overlap with cool spots. In that case, the magnetic enhancement can still be detected, but the recovered field strength is significantly underestimated, illustrating the common difficulty \citep[e.g.][]{rosen:2012} of probing starspot interiors with optical atomic spectral lines.

\section{Discussion}

\begin{figure}
\centering
\figps{\hsize}{0}{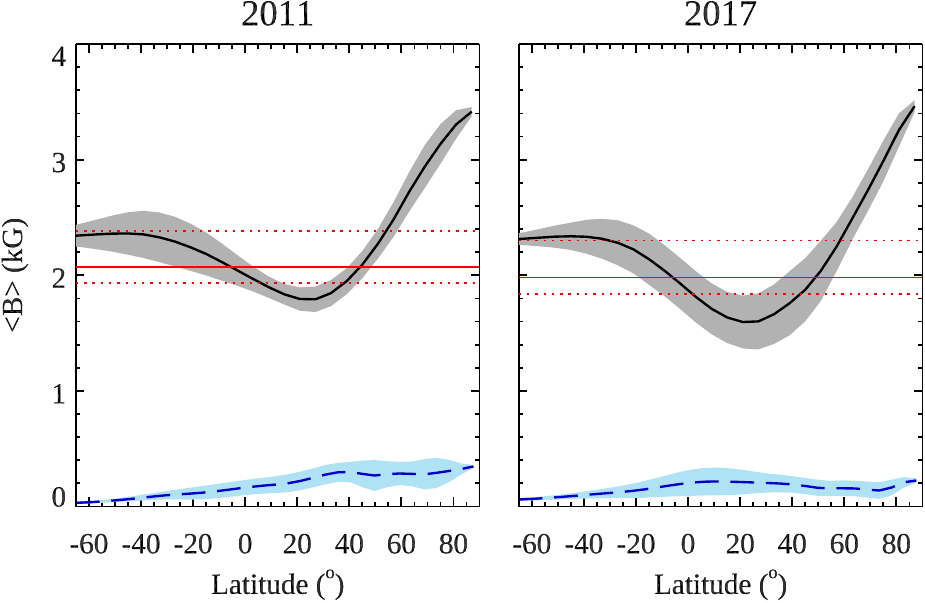}
\caption{Latitudinal dependence of the small-scale magnetic field strength for 2011 (left) and 2017 (right) epochs. The solid curves correspond to the longitudinally-averaged field strength value, the grey background indicates one standard deviation. The dashed lines on the blue background show the similar average for the global field according to ZDI results by \citet{lehtinen:2022}. The red horizontal lines show the spatially unresolved mean field strength (solid line) and associated uncertainties (dotted lines) derived by \citet{kochukhov:2020}.}
\label{fig:lat}
\end{figure}

A prominent latitudinal variation of the small-scale field strength uncovered by our analysis is illustrated in Fig.~\ref{fig:lat}. This plot also shows the unresolved field strength derived from the average spectra \citep{kochukhov:2020} and the latitudinal dependence of the global field strength from ZDI maps \citep{lehtinen:2022}. Both the present study and the spatially-unresolved analysis by \citet{kochukhov:2020} indicate that in 2017 the field on \lqh\ was slightly weaker compared to 2011. The average values from our DI field strength maps, 2.18 and 2.06 kG for the 2011 and 2017 data sets respectively, correspond to intermediate latitudes. The field drops to less than 2 kG around the equator and reaches nearly 3.5 kG at the visible rotational pole. We do not confirm the suggestion by \citet{saar:1992a,saar:1994b} that the small-scale field distribution on \lqh\ exhibits a significant non-axisymmetric component. Unlike these preliminary studies, our DI analysis yields a surface-averaged field intensity broadly in agreement with all previous studies, including the K-band measurement by \citet{saar:1996a} obtained with a completely different set of observational data and diagnostic lines. This suggests that our investigation provides the first reliable map of the small-scale magnetic field on an active late-type star.

The global magnetic field maps of \lqh\ were reconstructed by \citet{lehtinen:2022}, who applied ZDI to mean circular polarisation profiles derived from the same observations as analysed here. There is little correspondence between the global and small-scale field structures, apart from the tendency of both to strengthen towards the visible rotational pole in the 2011 data set. The average global field strength of \lqh\ is 169 and 157~G for the 2011 and 2017 epochs, respectively. Thus, the large-scale field is about 13 times weaker than the total field. Interestingly, both field components appear slightly weaker in the later set of observations.

The temperature maps of \lqh\ reconstructed in this study are qualitatively similar to the distributions recovered in other DI analyses which relied on interpretation of individual spectral lines \citep{flores-soriano:2017,cole-kodikara:2019}. Similar to those studies, we did not find a persistent cool polar cap. This contrasts our findings with those by \citet{lehtinen:2022}, who recovered a dark polar cap in the inversions based on mean (least-squares deconvolved) line profiles derived from the same HARSPpol spectra as studied here. A detailed discussion of this discrepancy is beyond the scope of this paper but it can be connected to simplifying line formation approximations employed in the DI modelling based on mean profiles.

One caveat of our work is that the \ion{Fe}{i} lines employed for simultaneous magnetic and starspot mapping are relatively strong, resulting in an increased impact of systematic errors and reduced sensitivity to temperature inhomogeneities. Ideally, the \ion{Fe}{i} lines studied here should be combined with weaker temperature-sensitive lines to improve reliability of starspot mapping. These limitations notwithstanding, the absence of spatial correlation between small-scale magnetic field and cool spots is noteworthy. This result suggests that cool spots on \lqh\ are either not associated with a particularly strong field or that this strong-field component cannot be resolved using optical data.

The uniformity of our small-scale magnetic field distribution suggests that this magnetic field component is unlikely to distort the reconstruction of spot geometries in non-magnetic DI. On the other hand, the total magnetic field of \lqh\ induces substantial line-dependent intensification that cannot be accounted for by non-magnetic inversions relying exclusively on theoretical or astrophysical line parameters validated using spectra of inactive stars. Empirical oscillator strength corrections might be necessary in the conventional DI of very active stars such as \lqh.

A qualitative picture of the overall surface magnetic morphology of \lqh\ is that of a highly-structured, intermittent magnetic field. A smoothed unsigned version of this magnetic topology corresponds to the field strength distribution derived here, whereas the vector averaging (affected by a massive bias due to cancellation of opposite polarities) results in much weaker global field maps inferred by ZDI studies. A joint self-consistent approach to modelling both spatial components of this multi-scale stellar magnetic field configuration is yet to be developed.

\begin{acknowledgements}
O.K. acknowledges support by the Swedish Research Council (grant agreements no. 2019-03548 and 2023-03667), the Swedish National Space Agency, and the Royal Swedish Academy of Sciences.
The work of T.H. was supported by the Research Council of Finland (project SOLSTICE, decision No. 324161).
\end{acknowledgements}


\onecolumn 

\begin{appendix}

\section{Test of DI using simulated observations}

\begin{figure*}[!h]
\centering
\figps{0.42\hsize}{0}{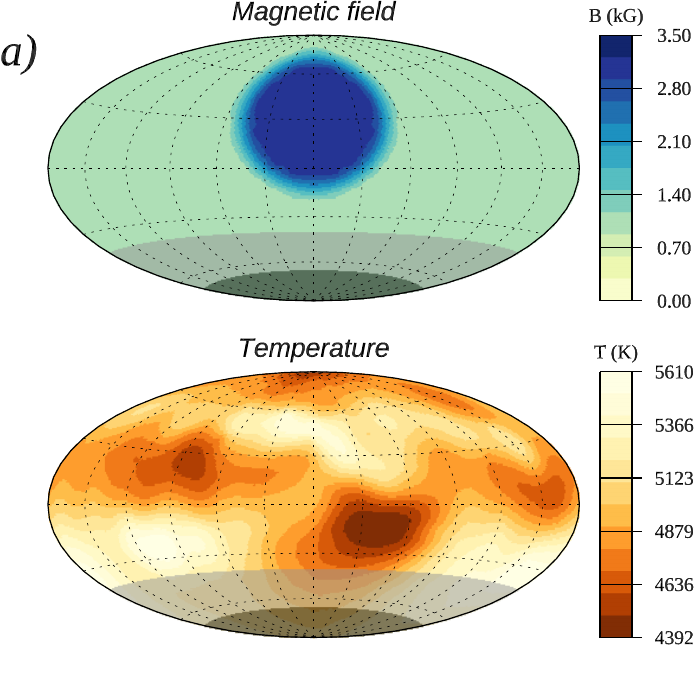}\hspace*{3mm}
\figps{0.42\hsize}{0}{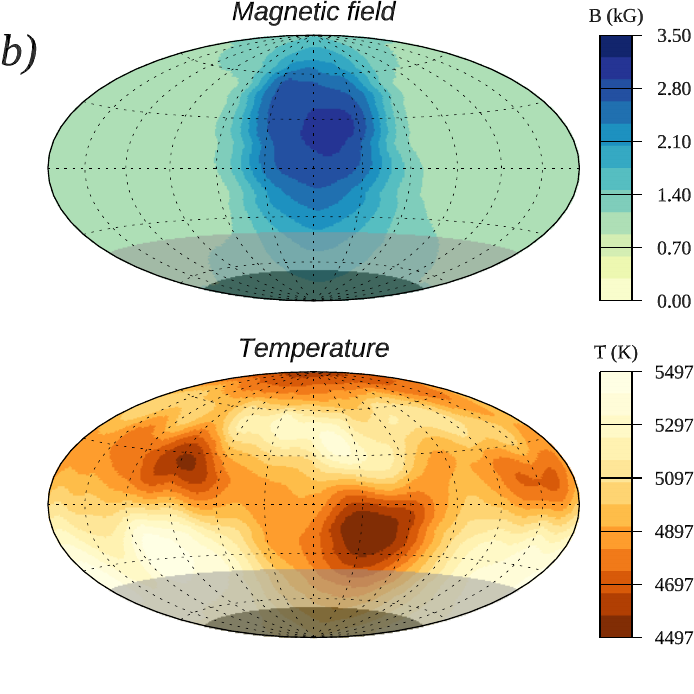}\\
\figps{0.5\hsize}{0}{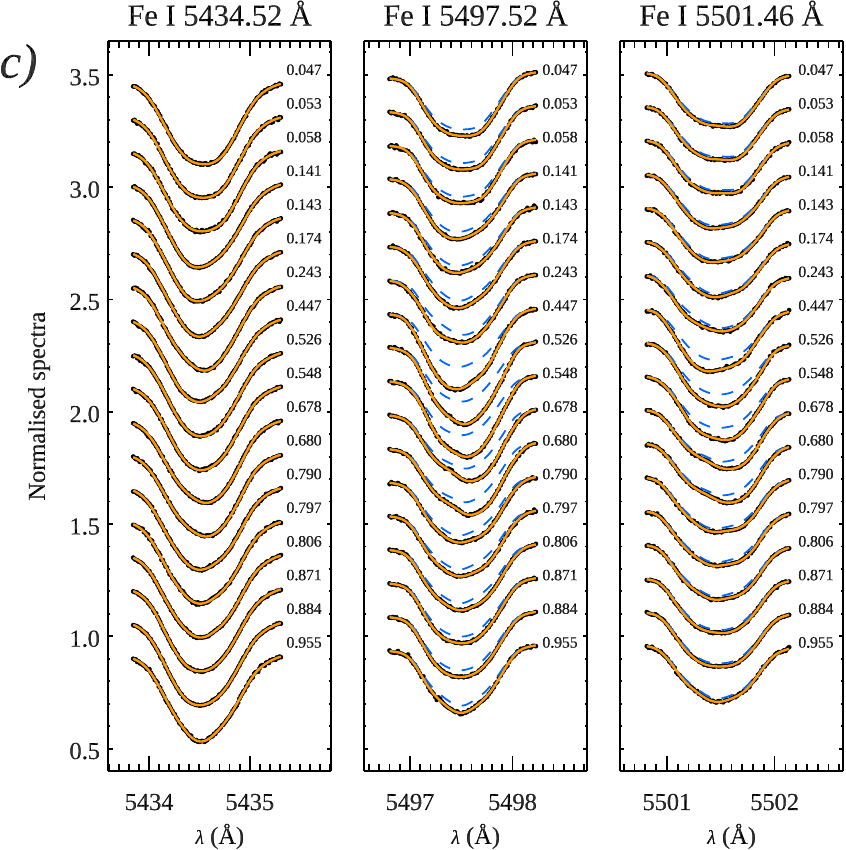}
\caption{Comparison of the input (a) and reconstructed (b) magnetic field and temperature distribution, together with the corresponding line profile fit (c) for the inversion test described in Sect.~\ref{sec:test}.}
\label{fig:test}
\end{figure*}

\end{appendix}

\end{document}